\documentclass[journal=jacsat,manuscript=article]{achemso}

\usepackage[version=3]{mhchem} 
\usepackage[labelfont=bf,textfont=normalfont]{caption}

\author{Bernardo S. Dias}
\affiliation[UvA]
{Van der Waals-Zeeman Institute, Institute of Physics, University of Amsterdam, Amsterdam, 1098 XH, the Netherlands}

\author{Reynolds Dziobek-Garrett}
\affiliation[ARCNL]
{Advanced Research Center for Nanolithography, Science Park 106, 1098 XG Amsterdam, the Netherlands}
\alsoaffiliation[UvA]
{Van der Waals-Zeeman Institute, Institute of Physics, University of Amsterdam, Amsterdam, 1098 XH, the Netherlands}

\author{Gabriella Mentasti}
\affiliation[UvA]
{Van der Waals-Zeeman Institute, Institute of Physics, University of Amsterdam, Amsterdam, 1098 XH, the Netherlands}

\author{Abhishek Gupta}
\affiliation[UvA]
{Van der Waals-Zeeman Institute, Institute of Physics, University of Amsterdam, Amsterdam, 1098 XH, the Netherlands}

\author{Alexander Lambertz}
\affiliation[AMOLF]
{Center for Nanophotonics, AMOLF, 1098 XG Amsterdam, The Netherlands}
\alsoaffiliation[UvA]
{Van der Waals-Zeeman Institute, Institute of Physics, University of Amsterdam, Amsterdam, 1098 XH, the Netherlands}

\author{Esther Alarcón-Lladó}
\affiliation[vtH]
{van ‘t Hoff Institute for Molecular Sciences, Universiteit van Amsterdam, Amsterdam, 1098 XH, the Netherlands}
\alsoaffiliation[AMOLF]
{Center for Nanophotonics, AMOLF, 1098 XG Amsterdam, The Netherlands}

\author{Peter Schall}
\affiliation[UvA]
{Van der Waals-Zeeman Institute, Institute of Physics, University of Amsterdam, Amsterdam, 1098 XH, the Netherlands}

\author{Roland Bliem}
\affiliation[ARCNL]
{Advanced Research Center for Nanolithography, Science Park 106, 1098 XG Amsterdam, the Netherlands}
\alsoaffiliation[UvA]
{Van der Waals-Zeeman Institute, Institute of Physics, University of Amsterdam, Amsterdam, 1098 XH, the Netherlands}

\author{Jorik van de Groep}
\affiliation[UvA]
{Van der Waals-Zeeman Institute, Institute of Physics, University of Amsterdam, Amsterdam, 1098 XH, the Netherlands}

\email{j.vandegroep@uva.nl}

\title[An \textsf{achemso} demo]
  {Large-Area Deterministic Stamping of 2D Materials on Arbitrarily Patterned Surfaces}

\abbreviations{IR,NMR,UV}

\begin{document}







\begin{abstract}
2D materials and their monolayers have attracted widespread interest by virtue of their unique electronic and optical properties. In addition to their remarkable physical characteristics, their atomically thin nature enables their integration in ultra-compact photonic and electronic devices, with potential for dynamic tunability via strain, charge carrier modulation or heterostructure engineering. While early research relied on micrometer-scale mechanically exfoliated flakes, recent advances—particularly gold-assisted exfoliation of transition metal dichalcogenides (TMDCs)—have enabled the preparation of high-quality, large-area monolayers, opening new opportunities for scalable device integration. For the field of nanophotonics in particular, the ability to transfer large-area 2D materials onto both flat and patterned substrates is essential for the development of functional devices. However, existing transfer techniques are often limited in scalability, compatibility with structured surfaces, or preservation of material quality. Here, we present a versatile and reliable transfer method of large-area monolayers and hBN/monolayer heterostructures onto both flat and nanostructured substrates. Our approach, based on the physical properties of low-density polyethylene, preserves the intrinsic optical quality of the materials and is compatible with a variety of device architectures. We demonstrate its applicability by fabricating devices that modulate the photoluminescence of TMDC monolayers through the manipulation of the photonic environment, strain or electrical gating. We further demonstrate the fabrication of van der Waals heterostructures using the same method. By enabling clean transfer of a wide range of monolayers and heterostructures, this technique offers a practical pathway for the development of next-generation optoelectronic platforms with improved functionality, scalability, and tunability.

\end{abstract}

\section{Introduction}

The development of 2D van der Waals (vdW) material fabrication and exfoliation procedures have  enabled unprecedented control of their layered structure, leading to breakthroughs in both the fundamental study of 2D material physics and the development of novel devices to manipulate charge carriers\cite{2DMatElectronicsReview}, light\cite{2DMatPhotonicsReview}, or spin waves\cite{2DMatSpinWavesReview} at the nanoscale. The vast range of 2D materials available\cite{2DMaterialList} has unveiled remarkable new material properties including large carrier mobility\cite{CarrierMobility2D}, bandgap tunability\cite{ BandgapEngineering2D}, and high thermal conductivity \cite{ThermalConductivity2D}. Moreover, the ability to stack different materials in heterostructures to explore new phenomena including Moiré physics\cite{2dMatMoireReview}, topological effects\cite{2DMatReviewTopological}, and interlayer excitons\cite{ModulationInterlayerExcitons}, further expands the applicability of vdW materials. 

In addition to the unique intrinsic properties of 2D materials and their heterostructures, engineered interactions with the underlying substrate can be leveraged to further tune the material's properties. For example, monolayer transition metal dichalcogenides (TMDCs) in contact with insulating materials support strongly bound excitons, which can be observed through bright photoluminescence (PL), while contact with metals can quench the emission\cite{PLQuenching, PLQuenching2}. Alternatively, (nano)patterned substrates have been employed to induce strain in monolayer TMDCs, leading to single photon emission from defect states\cite{StrainDefectEngineering, StrainEngineering2}, and the integration of such monolayers with optical metasurfaces\cite{MetasurfaceReview} has resulted in the observation of strong light-matter interactions, such as exciton polariton and plexitonic states \cite{ExcitonPolaritonTransport, Plexcitons}. The strong and tunable excitonic light-matter interaction has already enabled novel applications such as optical modulators\cite{ModulatorPlasmonicGrating, TomsModulator}, beam steering\cite{LudovicasBeamSteering}, and photodetectors\cite{PhotodetectorMoS2}. 
These exciting advances highlight the importance of the development of techniques that offer controllable stacking of these thin layers and integrate them with a wide variety of substrates.

While initial 2D material-based devices have shown very promising demonstrations that highlight the unique capabilities of these materials, they largely rely on small-area, mechanically exfoliated flakes. In this regard, several techniques have been developed to perform dry-transfer of these small flakes, relying on polymers such as PDMS\cite{PDMS_Stamping}, PVC\cite{PVC_Stamping} or PPC\cite{PPCStamping}, or Si$_3$N$_4$ membranes\cite{SiliconNitrideStamping}. Additional methods that directly combine these monolayers with hBN are readily proposed, with the notable example of the hot pickup technique, which  capitalizes on the stronger 2D material vdW attraction to pick up monolayers with the precise dimensions of the top hBN flake \cite{vdW_Pickup}. 

Recently, the development of the gold-assisted exfoliation (GAE) technique has marked a crucial advance in 2D material handling by enabling the exfoliation of TMDC monolayers up to centimeter scale, with material qualities comparable to mechanically exfoliated layers \cite{AuExfol_Science, AuExfol, AuExfol2}. At the same time, transferring these large and high-quality monolayers for the fabrication of large-area nanophotonic and electronic devices remains a critical challenge. In addition to the size limitation, existing transfer methods crucially depend on the reliable adhesion of the monolayer to the target substrate, which makes the transfer onto patterned surfaces very challenging. To mitigate this dependence, other methods have been devised where a small sacrificial layer of the polymer is transferred along with the heterostructure onto a flat surface, which is removed afterwards using a solvent \cite{ReviewTransferMethods}. Despite these advances, the ability to transfer large-area 2D materials onto arbitrarily patterned substrates with minimal surface adhesion remains a major outstanding challenge. At the same time, such ability would open a vast range of opportunities where high-quality 2D materials are integrated in large-area devices such as high-Q metasurfaces, array-based photodetectors, and flexible optoelectronic platforms.

Here, we demonstrate a simple method that allows reliable transfer of both large-area monolayers and hBN/monolayer heterostructures to patterned or unpatterned substrates, ranging from flat to high-aspect ratio and low adhesion patterned interfaces. In this process, we use low-density polyethylene (LDPE), an inexpensive and readily available polymer with advantageous properties such as low melting temperature and tunable surface energy \cite{LDPE_Properties, LDPE_Plasma}. 
By assembling and characterizing devices such as angularly selective metasurfaces, strain-engineered emitters, and electrically gated TMDC monolayers, we show how the technique facilitates the development of advanced 2D optoelectronic platforms. We also highlight how this transfer method enables the exploration of 2D material physics such as interlayer exciton formation over large areas, offering new possibilities for further fundamental studies of these materials.  

\section{Transfer of 2D material monolayers}

\begin{figure}[h]
    \centering
    \includegraphics[width=\linewidth]{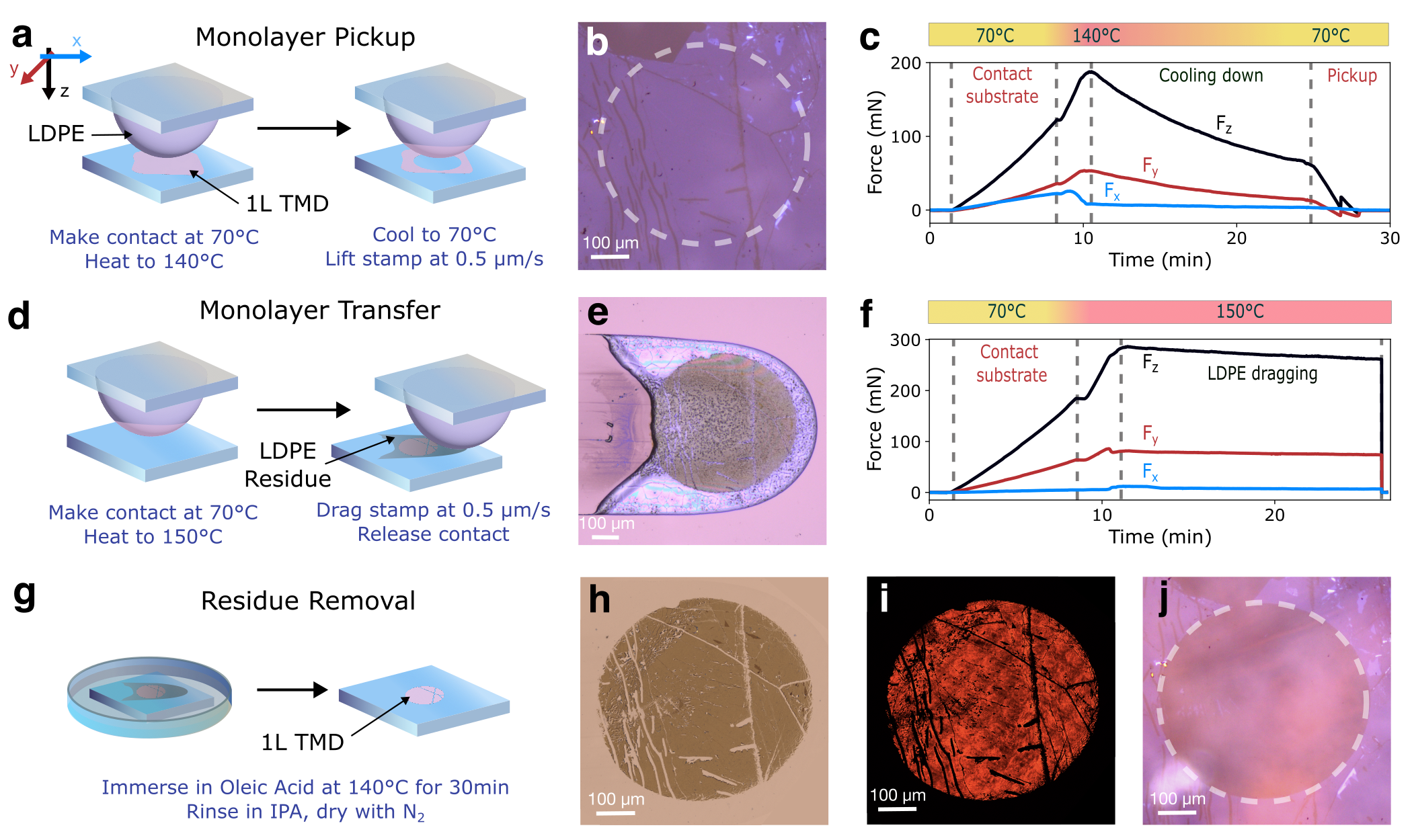}
    \caption{\textbf{Transfer method for large-area monolayers using the LDPE-based stamp. (a)} Procedure for monolayer pickup. \textbf{(b)} Bright field image of the WS$_2$ monolayer after GAE, before pickup. The cracks in the monolayer originate from the exfoliated bulk crystal. \textbf{(c)} Normal and in-plane force measurements during the pickup procedure. \textbf{(d)} Procedure for monolayer transfer. \textbf{(e)} Monolayer and LDPE residue after transfer to the target substrate. \textbf{(f)} Normal and in-plane force measurements during the transfer of the monolayer. \textbf{(g)} Procedure for LDPE residue removal. Bright field \textbf{(h)} and wide field PL \textbf{(i)} image of the transferred WS$_2$ monolayer. The grid-like structure with darker regions are stitching artifacts resulting from the merger of multiple high resolution microscope images into a large-area overview. \textbf{(j)} Bright field image of the initial substrate after transfer, highlighting the original transferred area.}
    \label{fig:1LTransfer}
\end{figure}

To analyze the proposed method without influence of substrate effects, we demonstrate the transfer of a GAE large-area WS$_2$ monolayer from a SiO$_2$ substrate to another target SiO$_2$ substrate (Fig. \ref{fig:1LTransfer}). We break the procedure into three discrete steps: the pickup, the transfer, and stamp residue removal.


The transfer method relies on the fabrication of a stamp composed of a heat-resistant superglue half-sphere covered by LDPE cling film (Fig. \ref{fig:1LTransfer}a, see SI section 1 for details on the fabrication of the stamp), which enables precise and selective pickup of 2D materials without contaminating other regions of the sample\cite{PVCDomeStamp, PVC_Stamping}. The stamp is placed in a plasma chamber for 2 minutes at 100~W under air plasma, which increases the adhesion of the stamp and reduces polymer residue after the transfer, as commonly reported for PDMS\cite{PDMSResiduePlasma} and PVC\cite{PVC_Stamping} stamping procedures. 
To pick up a 2D material monolayer, the stamp is placed on a motorized $xyz$-micrometer precision stage, and approaches the GAE monolayer (Fig. \ref{fig:1LTransfer}b) at 70°C with a speed of 0.5 \textmu m/s to make slow contact. We use force sensors in the stage to measure the forces involved during the stamping procedure (see SI section 2) both in the plane of the 2D material ($F_x$ and $F_y$) and perpendicular to it ($F_z$, Fig. \ref{fig:1LTransfer}c).
This not only offers better control and repeatability of the transfer, it also provides crucial information on the contact and friction dynamics throughout the process. During initial contact with the substrate, the normal force ($F_z$) increases to 120 mN, at which point we stop the approach. The in-plane force components also increase slightly due to a small tilt of the glass slide and coupling between $F_z$ and $F_y$ caused by the stamping tool geometry (see SI section 2). After the stamp has made contact with the monolayer, the system is heated to 140°C inducing a phase transition in which the LDPE melts, achieving conformal contact with the monolayer. This crucial step gives strong adhesion of the monolayer to the stamp and underpins the near-unity yield in the pickup process. As a result of thermal expansion during the phase transition, we see a rapid increase in $F_z$ (Fig. \ref{fig:1LTransfer}c). The system is then cooled down to 70°C, leading to LDPE solidification and a decrease in forces due to thermal contraction. We can then pick up the monolayer with a vertical motion at 0.5 \textmu m/s. Here, the spikes in the associated force curves indicate some intermittency in the detachment of the polymer due to adhesive forces\cite{Dan2023}.

Second, we discuss the monolayer transfer, in this example to a new quartz substrate, as illustrated in Fig.~\ref{fig:1LTransfer}d-f. We make contact with the substrate with the same parameters as for the pickup (0.5 \textmu m/s, 70°C), with Fig. \ref{fig:1LTransfer}f showing an initial force curve similar to Fig. \ref{fig:1LTransfer}c. The system is then heated to 150°C (rather than 140°C), further melting the LDPE and considerably reducing its viscosity compared to at 140°C\cite{LDPE_Properties}. Next, we slide the stamp parallel to the substrate interface along the $y$-axis away from the original contact area, leaving behind the monolayer covered by a sacrificial layer of LDPE (Fig. \ref{fig:1LTransfer}e). During the onset of dragging, we observe the transition from static to dynamic friction, marked by an oscillatory behavior in $F_y$ (Fig. \ref{fig:1LTransfer}f). The slow decrease in $F_z$, indicates possible mass loss due to the transfer of LDPE to the substrate (Fig.\ref{fig:1LTransfer}e). Finally, when the stamp is no longer in contact with the 2D material, we safely lift it from the substrate, which then slowly cools to room temperature. 

Third, we remove the residual LDPE from the top side of the monolayer (Fig. \ref{fig:1LTransfer}e) by placing the sample in oleic acid for 30 minutes at 140°C, rinsing in isopropanol (IPA), and drying with N$_2$ (Fig. \ref{fig:1LTransfer}g). Our transfer technique preserves the macro-scale topography of the monolayer while inducing minimal additional cracks throughout the process (compare Fig. \ref{fig:1LTransfer}b,h), and the material sustains its excitonic PL that is characteristic for WS$_{2}$ monolayers (Fig. \ref{fig:1LTransfer}i). After transfer, the original substrate is left with a clean area without monolayer (Fig. \ref{fig:1LTransfer}j).  We emphasize that while Fig. \ref{fig:1LTransfer} shows the transfer of a monolayer $\sim$0.5~mm in diameter, the size is only limited by the geometry of the monolayer after the GAE process and the controllable contact area between the stamp and the monolayer, which enables both smaller and larger areas to be transferred.\\ 

\begin{figure}[h]
    \centering
    \includegraphics[width=.6\linewidth]{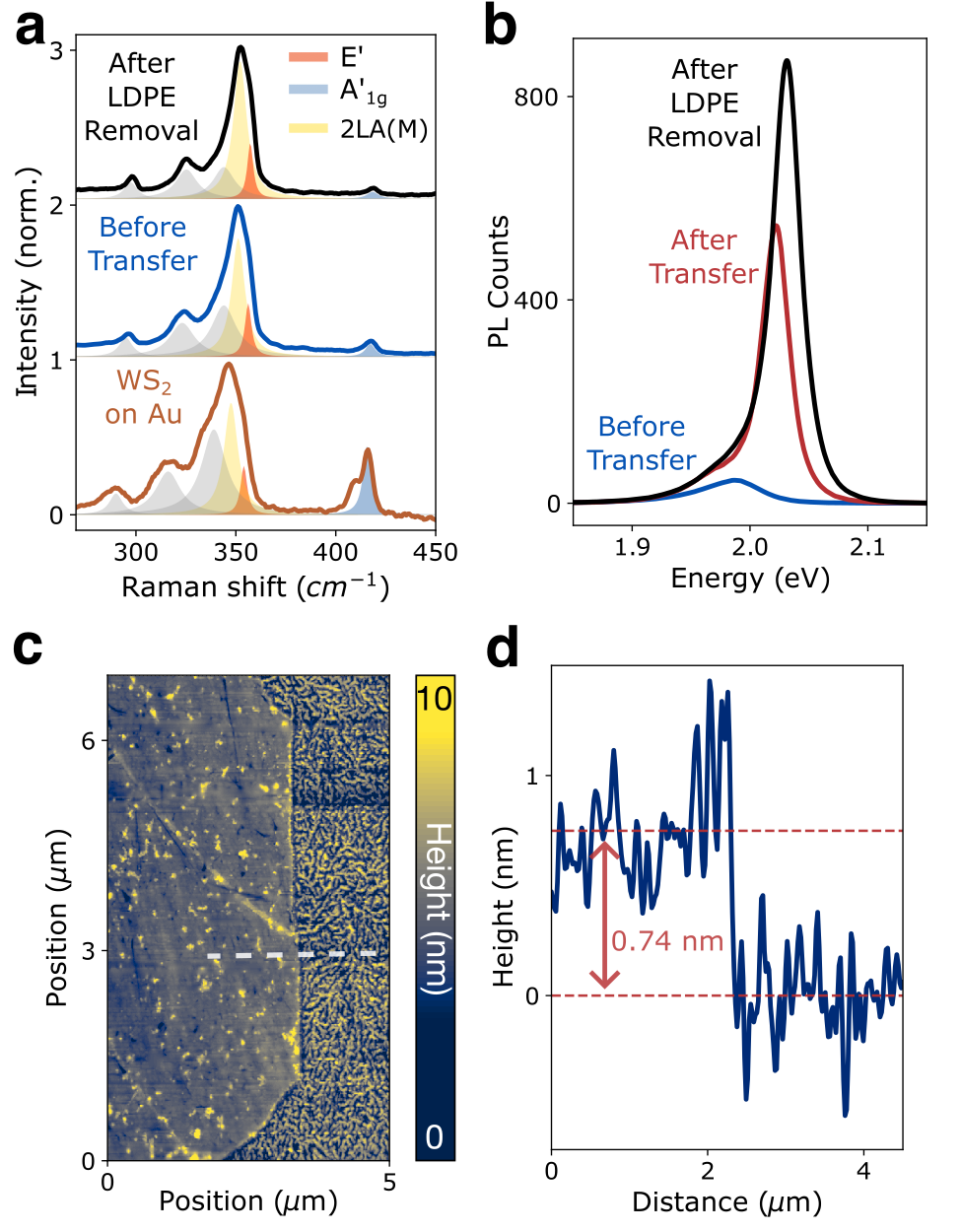}
    \caption{\textbf{Characterization of the monolayer before and after transfer. (a)} Raman and \textbf{(b)} PL spectra of the monolayer before and after transfer, as well as after LDPE residue removal. \textbf{(c)} AFM map and \textbf{(d)} averaged line scan of the monolayer, showing the thickness after transfer. The white dashed line in (c) highlights the central location where the line scan in (d) is taken.}
    \label{fig:1LTransfer_Characterization}
\end{figure}

By using the same material (SiO$_2$) for both the source and target substrate, we can directly monitor the impact of the transfer on the material quality. First, to assess possible transfer-induced strain, we measure Raman spectra of the monolayer during multiple stages of the procedure (Fig. \ref{fig:1LTransfer_Characterization}a). We observe a considerable redshift of the $E^1_{2g}$ peak when WS$_2$ is exfoliated from the bulk crystal during the initial GAE step (354 cm$^{-1}$, measured on Au), in agreement with recent literature  that argues that GAE is mediated by biaxial strain induced in the monolayer, decreasing the bond strength between the first and second layers\cite{AuExfolStrain}. Comparing the same $E^1_{2g}$ peak for monolayer WS$_2$ on quartz before pickup (356 cm$^{-1}$) and after transfer and LDPE cleaning (357 cm$^{-1}$) a smaller shift is observed, which we argue is mostly caused by changes in carrier doping rather than strain (see SI section 3 for a detailed discussion, Figs. S4 a,b). As such, we conclude that the transfer process does not induce a significant strain in the monolayer. 

Two crucial observables that characterize the quality of monolayer TMDCs for applications in photonic devices are the PL linewidth and intensity. Analogous to the Raman analysis, we study the excitonic PL spectra at different stages of the transfer process (Fig. \ref{fig:1LTransfer_Characterization}b). Compared to the as-exfoliated film (blue), the LDPE transfer induces a marked improvement in the PL by transitioning from a spectrum dominated by emission from trions to neutral excitons. Here, the exciton linewidth decreases from 38 to 26\,meV, combined with a strong enhancement of the PL emission intensity (red). These LDPE-induced changes are unexpected and are not previously reported in the literature. Combining analysis of both Raman and PL data, we propose that the PL spectrum changes are induced by a variation in the doping level of the monolayer (see SI section 3 for a detailed discussion). This process may be induced by electron transfer from the surface functionals created by the plasma process in LDPE. Lastly, the subsequent removal of the LDPE residue with oleic acid maintains the narrow linewidth and further increases the emission intensity, reaching a final enhancement over 19 times. Previous work using oleic acid on TMDCs suggests that the enhanced PL quantum yield results from: (i) passivation of sulfur vacancies, reducing the density of trap states available; and (ii) protection of the monolayer from atmospheric reactants by the alkyl tail of the oleic acid encapsulating the surface \cite{OleicAcid1, OleicAcid2}.

Finally, to explore any polymer residue on the surface of the monolayer leftover from the transfer process, we perform atomic-force microscopy (AFM) scans on the monolayer after transfer (Figs. \ref{fig:1LTransfer_Characterization}c-d). A small scale residue is formed on the surface of the substrate, which can be attributed to the formation of an oleic acid monolayer, as previously observed\cite{OleicAcid1L_1, OleicAcid1L_2}. Interestingly, this residue is mostly absent on the monolayer, possibly due to surface energy differences between the substrate and the 2D material. Taking a line section and averaging over the neighboring pixels, we verify a height step of 0.74 nm, showing absence of a layered residue on top of the monolayer. Overall, we conclude that the presented method not only allows for the clean transfer of the monolayer but simultaneously improves the exciton emission from the TMDC samples.


\begin{figure}[h]
    \centering
    \includegraphics[width=\linewidth]{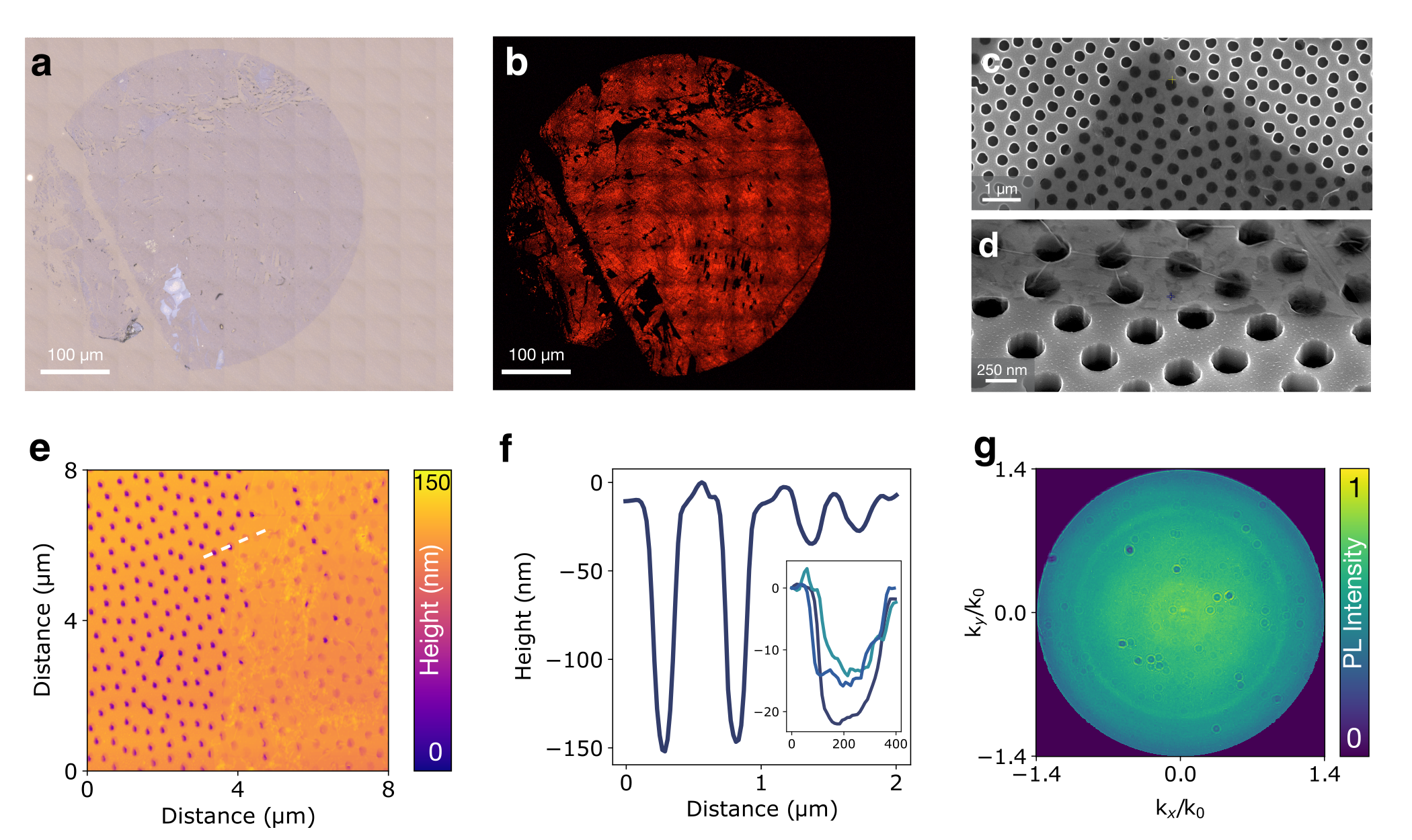}
    \caption{\textbf{Transfer of a WS$_2$ monolayer on a SiO$_2$/Si substrate with hyperuniform patterned nanoholes. (a)} Bright field and 
\textbf{(b)} wide field PL images of the transferred WS$_2$ monolayer. Top view \textbf{(c)} and tilted \textbf{(d)} SEM images of the monolayer conforming to the substrate, where monolayer draping over the holes is clearly visible. AFM  map \textbf{(e)} and cross section \textbf{(f)} of the monolayer on the patterned substrate. Inset: zoomed traces of several monolayer covered nanoholes showing a typical height variation of $\sim15$~nm above the holes. \textbf{(g)} Back focal plane (BFP) microscope image of the PL emission of the sample, showing a ring with emission enhancement at $k_{x}/k_{0}=1.016$ due to light scattering by the hyperuniform pattern.}
    \label{fig:1LonSiNanoholes}
\end{figure}

Besides flat surfaces, our proposed method also enables transfer to nanopatterned substrates, where the surface adhesion is smaller due to a reduced contact area. To illustrate this, we transfer a large WS$_2$ monolayer onto a SiO$_2$ (55\,nm) / Si substrate patterned with nanoholes (covering 25\% of the surface area) arranged in a hyperuniform pattern (Fig. \ref{fig:1LonSiNanoholes}). We obtain a 0.18 mm$^2$ covered area with uniform PL emission (Fig. \ref{fig:1LonSiNanoholes}a,b). Using scanning-electron microscopy (SEM), we confirm that the monolayer dresses the nanopatterned surface with the material covering the nanoholes without rupturing (Fig. \ref{fig:1LonSiNanoholes}c,d). To quantitatively assess the topography of the monolayer on top of the pattern, we perform AFM measurements and observe that the uncovered holes display a depth of approximately 150\,nm, while the WS$_2$-covered holes exhibit a very minor height profile due to draping of the monolayer over the hole (Fig.~\ref{fig:1LonSiNanoholes}e,f). 

The transfer of 2D materials onto such large-area nanopatterns opens new opportunities for studying the interaction of delocalized photonic resonances with material resonances such as excitons. In this example, by placing a WS$_2$ monolayer on top of a hyperuniform pattern, we demonstrate the ability of directional PL emission through modification of the photonic environment. An important characteristic of hyperuniform patterns is the suppression of the structure factor $S(k)$ at $k=0$, where $k$ is the in-plane wavevector of light. This effect manifests itself as an absence of long-range density fluctuations while simultaneously showing a disordered local arrangement that remains statistically uniform over large scales, without periodicity\cite{Hyperuniform}. This suppression results in a redistribution of scattered light intensity to larger wavevectors, which in this case lie beyond the light line in air. Using back focal plane (BFP) measurements with oil immersion, we measure a ring-like enhancement of the PL at $k_x / k_0$ = 1.016, corresponding to a wavevector of $k_x$ = 10.3 $\mu m ^{-1}$ (Fig. \ref{fig:1LonSiNanoholes}g). From SEM images we also extract $S(k)$ of the pattern (SI section 4), verifying excellent agreement between the emission enhancement in the BFP and the peak in the structure factor. Altogether, this example demonstrates how our LDPE transfer technique offers novel opportunities for the merger of large-area 2D monolayers and photonic metasurfaces with decreased surface adhesion. 


\section{Transfer of hBN/1L heterostructures}

\begin{figure}[h]
    \centering
    \includegraphics[width=\linewidth]{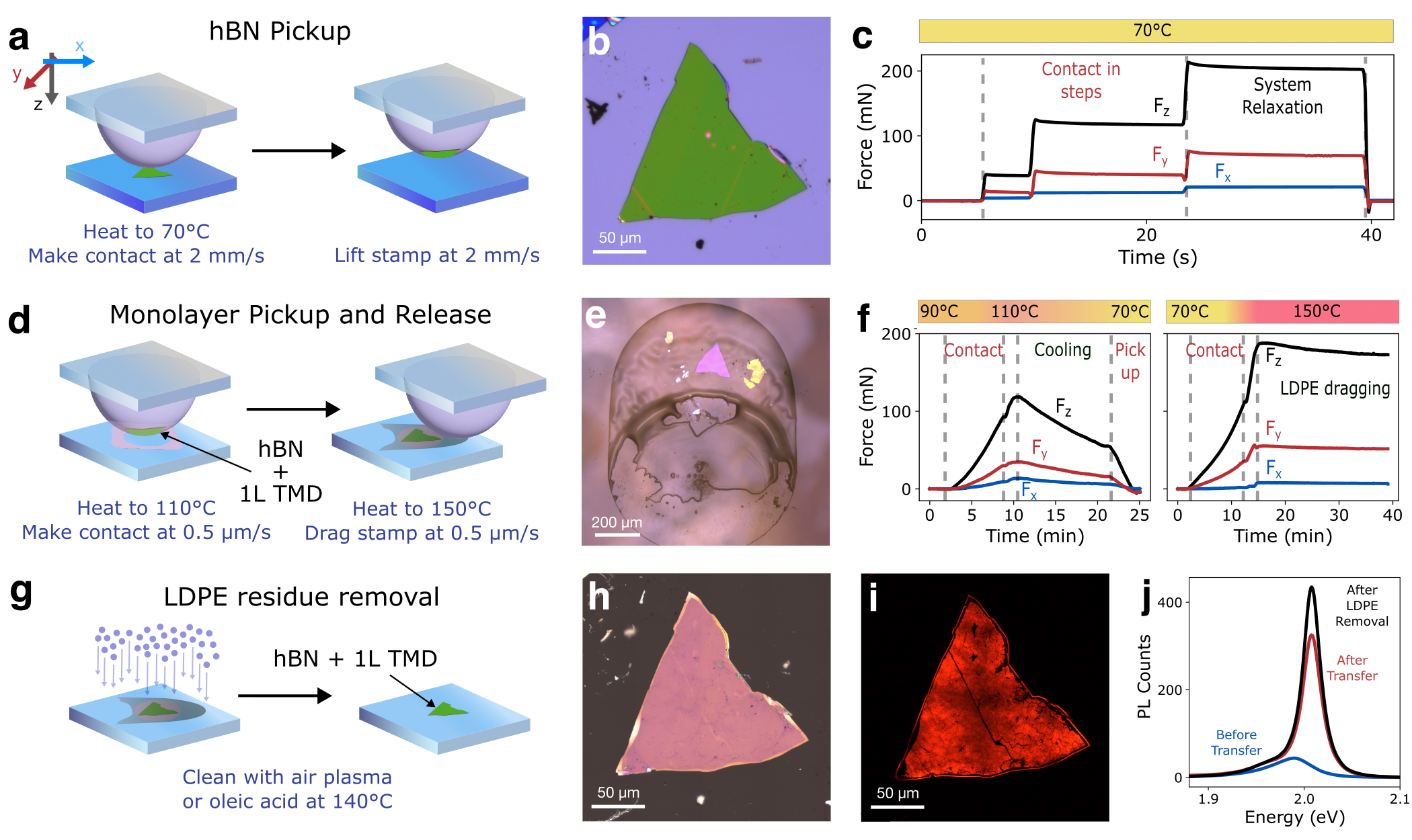}
    \caption{\textbf{Transfer of hBN/monolayer heterostructures. (a) }Procedure for hBN pickup. \textbf{(b)} Bright field image of the hBN flake before transfer. \textbf{(c)} Normal and in-plane force measurements during the pickup of an hBN flake. \textbf{(d)} Procedure for monolayer pickup with hBN and release on a target substrate. \textbf{(e)} Bright field microscope image of the flake after deposition, covered in LDPE residue. \textbf{(f)} Normal and in-plane force measurements during the pickup of the monolayer with hBN (left) and release of the structure on the final substrate (right). \textbf{(g)} Procedure for LDPE removal by air plasma treatment. \textbf{(h)} Bright field and wide field PL \textbf{(i)} images of the hBN flake after LDPE residue removal. \textbf{(j)} PL spectra before, after transfer, and after LDPE residue removal.}
    \label{fig:hBNWS2Transfer}
\end{figure}

One of the crucial features in 2D material research is the ability to stack different layers into heterostructures, enabling the assembly of novel materials that outperform the sum of their individual components. In particular, heterostructures combining hexagonal boron nitride (hBN) with TMDC or graphene monolayers have been employed extensively to provide atomically flat surfaces free of defects and charge traps. This improves the material's carrier mobility and preserves the intrinsic optical and electronic properties of the monolayers\cite{ReviewHeterostructures}. Here, we  demonstrate how our LDPE transfer technique can be used to fabricate large-area hBN (top) / 1L (bottom) heterostructures (Fig. \ref{fig:hBNWS2Transfer}). 

We pick up a mechanically exfoliated hBN flake from a Si substrate at 70°C using a plasma-activated LDPE stamp, by making contact in steps of 0.1 mm at 2 mm/s (Fig. \ref{fig:hBNWS2Transfer}a-b). While the step height remains constant, we observe an increase in the force exerted at each step, attributed to the larger contact area between the curved dome and the flat substrate (Fig. \ref{fig:hBNWS2Transfer}c). After each step, we allow the system to relax for a few seconds, during which the force decreases slightly, visible as a small exponential relaxation following the initial force spike. When the hBN flake is fully covered by the LDPE, we lift the stage at a speed of 2 mm/s. We find that this method of making contact and picking up at large speeds induces fewer cracks in the hBN flake, as also observed in ref.\cite{PVCDomeStamp}. 


For the subsequent pickup of a WS$_2$ monolayer with the stamp and hBN, we find that the relative humidity of the environment and the surface energy of the monolayer and carrier substrate play a significant role in the adhesion of the monolayer to the hBN flake. As such, we first cover the monolayer with a self-assembled monolayer (SAM) of 1-decanol (see SI section 1 for further information), which passivates the surface and reduces the interaction with atmospheric contaminants \cite{DodecanolCovering}, thereby enabling a clean monolayer pickup. The stamp with hBN is then put in contact with the monolayer at a temperature of 90°C and a speed of 0.5 \textmu m/s (Fig. \ref{fig:hBNWS2Transfer}d), leading to an increase in $F_z$ (Fig. \ref{fig:hBNWS2Transfer}f, left). After contact, the structure is heated to 110°C for 2 minutes, promoting the mobility of any air or water pockets trapped between the hBN and the monolayer and improving contact between layers. After cooling the stamp back to 70°C, the stamp is pulled upwards at a rate of 0.5 \textmu m/s. In this process, both the 1L underneath and outside the hBN flake in contact with the stamp is picked up. This presents a marked difference from the previously reported van der Waals pickup technique, where only the 1L in contact with hBN is picked up \cite{vdW_Pickup}. Crucially, neither the top of the monolayer nor the bottom of the hBN flake has been in contact with the LDPE, which results in a clean heterostructure interface. To transfer the heterostructure to the target substrate, we employ a similar routine as for the single 1L transfer, with a contact at 70°C at 0.5 \textmu m/s followed by heating up to 150°C to melt the LDPE. The stamp is then dragged at a speed of 0.5 \textmu m/s and pulled up once the heterostructure is no longer in contact with the stamp, resulting in the sample covered by LDPE (Figure \ref{fig:hBNWS2Transfer}e). Due to the similarity of the procedures, the force curve of this transfer process (Fig. \ref{fig:hBNWS2Transfer}f) closely mimics Fig. \ref{fig:1LTransfer}c,f. 


There are two different methods to remove the LDPE from the heterostructure. The first follows the procedure described above for monolayers, removing the LDPE with oleic acid at 140°C. However, for structures stamped on substrates with minimal adhesion or large aspect-ratio features, potential solvent intercalation between the surface pattern and the heterostructure may result in detachment of the heterostructure from the surface. The second method, which we employ here, is to remove the LDPE with soft air plasma cleaning (Fig. \ref{fig:hBNWS2Transfer}g). After this process, the hBN flake does not show visible damage (Fig.~\ref{fig:hBNWS2Transfer}h) and fully covers a continuous WS$_2$ monolayer, as visible in the wide-field PL image (WFPL, Fig.~\ref{fig:hBNWS2Transfer}i). To assess any structural damage to the hBN flake after plasma etching, we perform AFM measurements that show no significant vertical etching of the hBN, only a subtle slanting of the side walls of the flake (SI, section 5). We note that the difference in flake color between Figs. \ref{fig:hBNWS2Transfer}b,h is due to a difference in the substrate (Si and SiO$_2$ for b,h, respectively). To analyze the effect of the heterostructure transfer and subsequent plasma cleaning on the excitonic properties, we measure the PL of the structure (Fig. \ref{fig:hBNWS2Transfer}j). After the transfer process, the linewidth decreases from 34 to 21\,meV due to hBN passivation. Next, after the plasma cleaning step the exciton linewidth remains 21\,meV, demonstrating that the underlying monolayer is not affected by the etching process. In addition, we observe a ten-fold increase in the PL intensity as a result of the hBN passivation and potentially a small Purcell enhancement due to internal reflections and interference patterns inside the hBN. Altogether, Fig.~4 shows how the proposed transfer method also enables the direct transfer of large and high-quality hBN/TMD heterostructures. We emphasize that the size of the heterostructure is only limited by the size of the hBN flake -- which is mechanically exfoliatied -- and thus considerably smaller than the monolayer obtained through the GAE method. 

\begin{figure}[h]
    \centering
    \includegraphics[width=\linewidth]{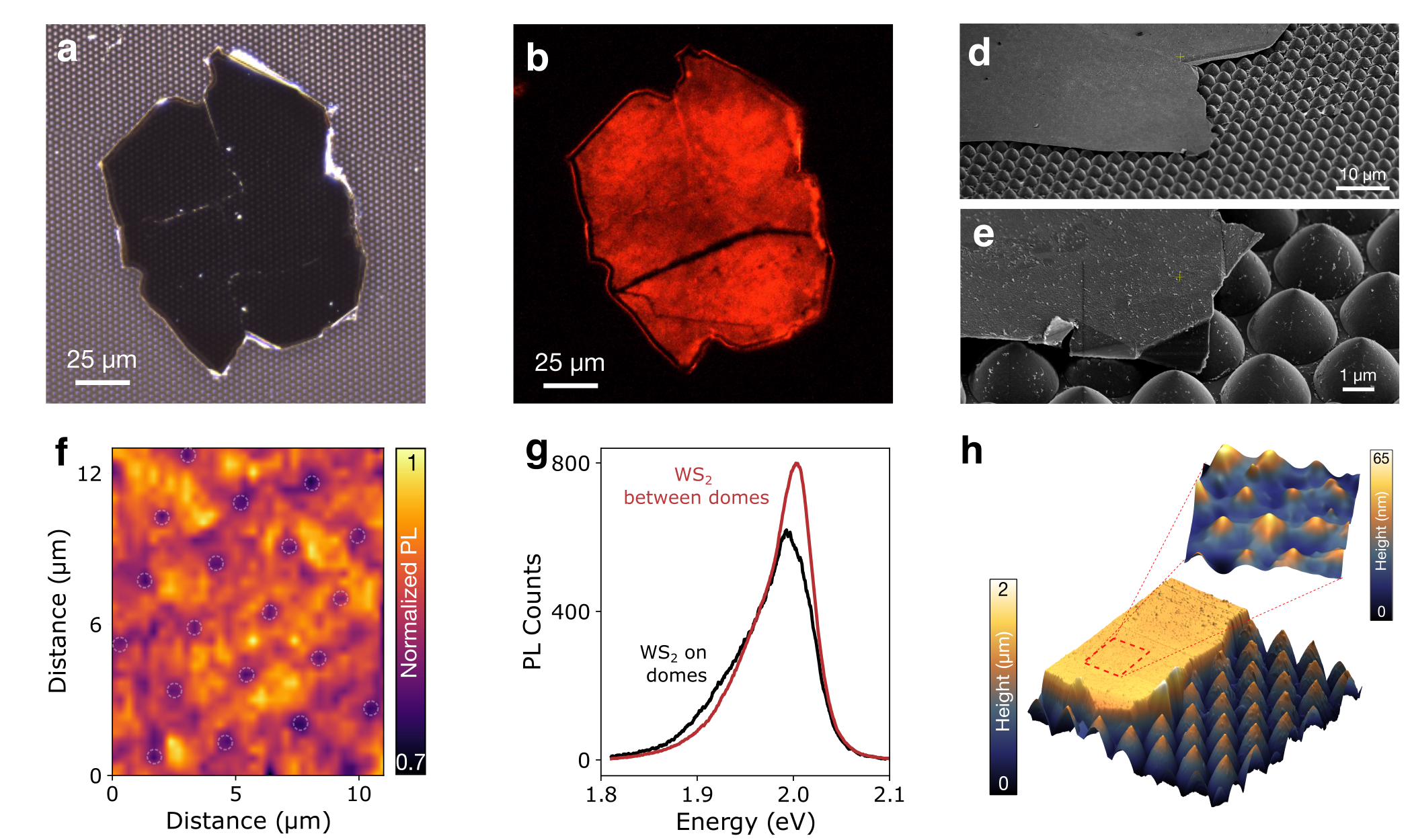}
    \caption{\textbf{Transfer of a hBN/WS$_2$ heterostructure onto a dome-patterned sapphire substrate.} \textbf{(a)} Dark field and \textbf{(b)} WFPL microscope images of the heterostructure on the domes after the transfer and plasma cleaning. \textbf{(d,e)} Tilted SEM images of the heterostructure on top of the sapphire domes. \textbf{(f)} PL map of the heterostructure, where each pixel represents the integrated PL intensity. The points of contact are highlighted with white dashed circles. \textbf{(g)} PL spectra of the structure on tip and in between the domes. \textbf{(h)} AFM measurement of the heterostructure on top of the domes. Inset: small-area measurement of top of the flake.}
    \label{fig:hBNWS2Transfer_Demonstration}
\end{figure}

To push the limits of the applicability of the presented method, we now explore the transfer onto high-aspect ratio structures. In particular, we transfer a 350\,nm hBN / 1L WS$_2$ heterostructure onto a sapphire substrate patterned with 1.8 \textmu m high domes (Fig. \ref{fig:hBNWS2Transfer_Demonstration}). After transfer and plasma cleaning, we place the substrate in a vacuum oven at 300°C for one hour to promote the degradation of any residual LDPE trapped in between the sapphire domes\cite{LDPE_Pyrolysis}. 

Contact between the heterostructure and the substrate is limited to the tips of the domes, resulting in exceptionally small points of contact. To assess the effect of the transfer on the heterostructure, we record both dark field and WFPL images showing that although the WS$_2$ monolayer is directly in contact with the domes, no signs of mechanical damage induced by the dragging process are observed (Fig. \ref{fig:hBNWS2Transfer_Demonstration}a,b). The stripe where there is no WS$_2$ under the hBN results from the exfoliation process, and was already present before transfer. Furthermore, SEM images of the heterostructure show that the flake lies suspended on top of the domes due to the high rigidity of hBN (Fig. \ref{fig:hBNWS2Transfer_Demonstration}d,e). 

 To assess the impact of strain, we measure a PL map and spectra of the heterostructure on the domes (Figure \ref{fig:hBNWS2Transfer_Demonstration}f,g). The points of contact are clearly visible due to increased strain at these locations, causing a local 30\% decrease of the PL emission. Comparing the PL spectrum of a bare WS$_2$ monolayer on a flat substrate (Fig. \ref{fig:1LTransfer_Characterization}b, before transfer) with spectra on and between the domes, we see significant exciton broadening, a redshift and a larger trion contribution. These effects are in agreement with previous observations, where tensile strain applied to monolayer WS$_2$ shows not only a redshift due to a reduced bandgap caused by lattice stretching \cite{WS2_BandgapStrain}, but also increased trion contributions \cite{WS2_Strain}. As expected, the PL spectrum on the domes shows a stronger redshift than in between domes, indicating that the tensile strain near the point contacts is larger. This interpretation is corroborated by an AFM scan of the structure (Fig. \ref{fig:hBNWS2Transfer_Demonstration}h). While at larger scale the hBN flake appears to lie flat on the domes, the inset demonstrates that the flake drapes about 65\,nm over the 3 \textmu m separation between points of contact, inducing large strain at this regions. 
 Overall, our results demonstrate that transfer of large-area hBN / monolayer heterostructures is possible even onto high aspect-ratio patterns, where the contact area between the materials and the substrate is extremely low, which is incompatible with most existing dry-transfer techniques. 





\section{Applications of the method}

\begin{figure}[h]
    \centering
    \includegraphics[width=.5\linewidth]{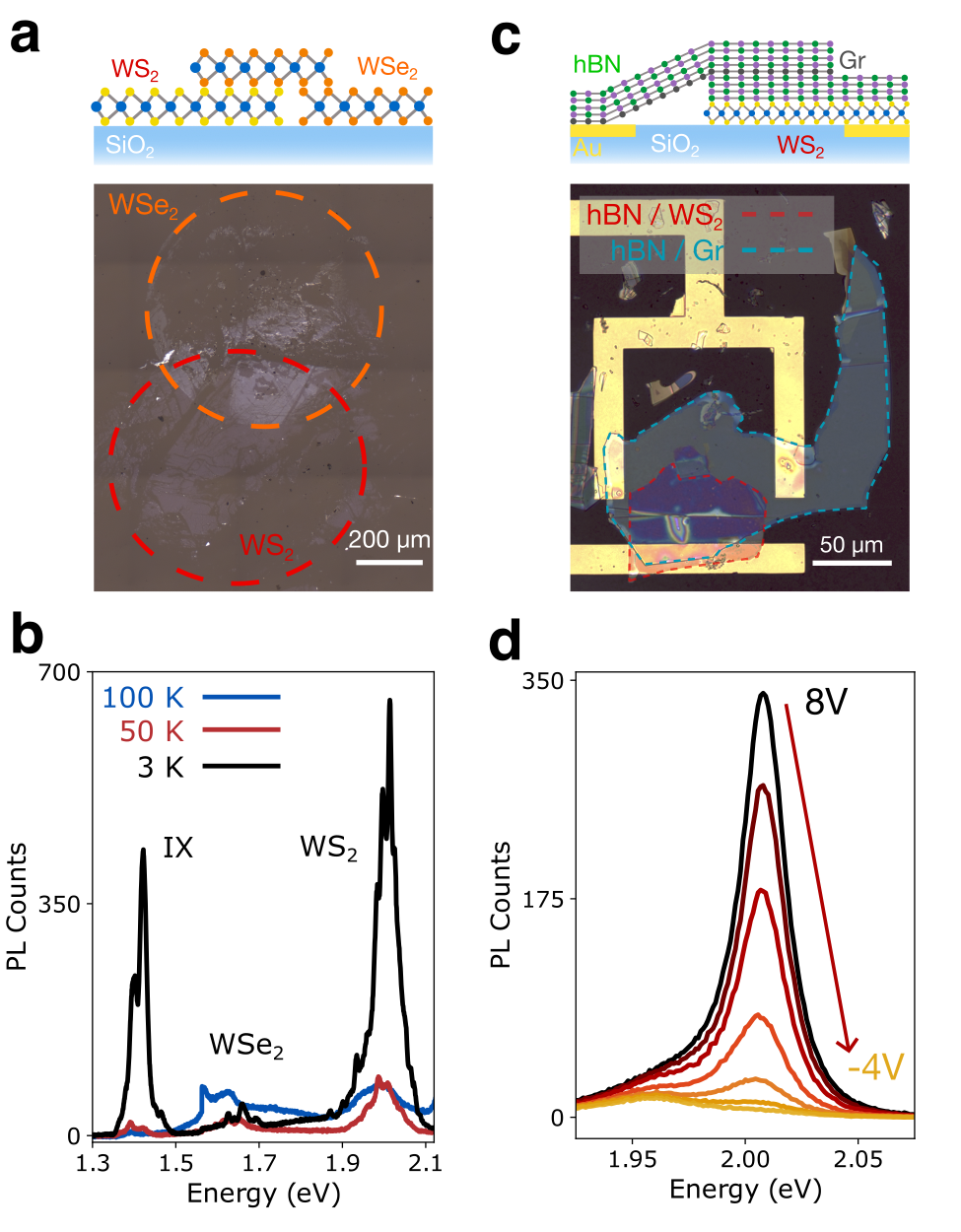}
    \caption{\textbf{Demonstration of  applications of the proposed transfer method. (a)} Bright field microscope image of the WS$_2$/WSe$_2$ heterostructure. \textbf{(b)} PL spectra at temperatures of 100 (blue), 50 (red), and 3 K (black). \textbf{(c)} Bright field microscope image of the hBN/graphene/hBN/WS$_2$ heterostructure stamped on Au contacts. \textbf{(d)} PL spectra at room temperature for different applied voltages.}
    \label{fig:Applications}
\end{figure}

The two transfer techniques presented enable the assembly of high-quality, large-area 2D material structures and their combination with patterned or unpatterned substrates, which  offers unique opportunities for techniques and device architectures which require large areas of continuous monolayer coverage. Here, we demonstrate two exciting possibilities of the proposed method (Fig. \ref{fig:Applications}): the fabrication of large-area bilayer heterostructures, and tunable optoelectronic devices with electronic control over the material's charge carrier density. 

First, we fabricate a large-area WS$_2$-WSe$_2$ heterostructure (Fig. \ref{fig:Applications}a). WS$_2$ is transferred on a SiO$_2$ substrate using the procedure outlined in Fig. \ref{fig:1LTransfer}, followed by a similar transfer of WSe$_2$. The heterostructure is then annealed in vacuum at 300°C to promote contact between the materials. The two layers are stamped with partial overlap only, to provide areas for the characterization of the heterostructure as well as the bare WS$_2$ and WSe$_2$ monolayers as a reference. To probe the excitonic behaviour of the heterostructure, we measure PL spectra at temperatures of 100, 50, and 3 K (Fig. \ref{fig:Applications}b). We observe the presence of the intralayer excitons of WS$_2$ and WSe$_2$ at 2 and 1.65\,eV, respectively, and a third resonance at 1.4\,eV (labeled $IX$), consistent with literature reports of interlayer excitons in WS$_2$-WSe$_2$ heterostructures \cite{WS2WSe2Heterostructure}. The three excitonic resonances demonstrate different behavior as function of temperature. While both the WS$_2$ intralayer exciton and the interlayer exciton increase their PL emission intensity at 3 K (as expected due to reduced exciton-phonon scattering), the WSe$_2$ intralayer exciton is quenched. For the heterobilayer, the electron of the photogenerated excitons in WSe$_2$ quickly transfers to the conduction band of WS$_2$, while the hole remains in the valence band of WSe$_2$. This spatial separation forms an interlayer exciton within a sub-picosecond timescale, which exhibits a long lifetime\cite{WS2WSe2Heterostructure}. This mechanism directly suppresses the PL from WSe$_2$ in the heterostructure as the electron transitions to the WS$_2$ layer before it can recombine radiatively. The observation of interlayer exciton emission is a clear indication of electronic coupling between the layers, and thus that our method provides clean interfaces in the heterostructure assembly.

Next, we assemble a more complex heterostructure device including contacts and electrical gates to modify the charge carrier density of the material, enabling control of the optical properties of the heterostructure (Fig. \ref{fig:Applications}c). Using the procedure of Fig. \ref{fig:hBNWS2Transfer}, fabrication of the device is achieved through a two-step process of transferring an hBN (100\,nm)/WS$_2$ heterostructure followed by an hBN (90\,nm)/graphene heterostructure. The WS$_2$ and the graphene connect to two different electrodes, and the middle hBN layer acts as an insulator, preventing short circuit below its breakdown voltage. The transfer of the hBN/graphene heterostructure highlights that the transfer method can also be applied to 2D monolayers other than TMDCs. Here, the high conductivity of the graphene enables its function as the gate electrode.\cite{PlasmonicModulator}. We probe the effect of electrical gating on the exciton by measuring the PL at different gate voltages (Fig. \ref{fig:Applications}d). The PL of the WS$_2$ monolayer can be fully quenched by inducing strong n-type doping at negative bias; we also observe PL enhancement as we approach 8\,V. This increase in the neutral exciton emission with increasing positive bias indicates the possibility of intrinsic n-doping of the WS$_2$ 1L, in agreement with previous results \cite{TomsModulator}. These demonstrations show the broad range of possibilities enabled by the LDPE transfer method, both for fundamental studies of large-area 2D material heterostructures as well as the development of novel optoelectronic devices.

\section{Discussion}

We present LDPE as a polymer carrier for 2D material transfer, for which polymers such as PDMS \cite{PDMS_Stamping}, PPC \cite{PPCStamping}, and PVC \cite{PVC_Stamping, PVCDomeStamp}, and others\cite{TransferMethodsReview} have also been used. Although these polymers have advantages for transferring specific 2D material layers, here we emphasize the particular attributes of LDPE that make it a suitable candidate for 2D material handling. Reliable 2D material transfer depends on precise tuning of the adhesion of the carrier polymer—high adhesion for pickup and low adhesion for release. Due to its 40--50\% crystallinity, LDPE undergoes a melting transition upon heating as the crystalline domains collapse\cite{LDPE_Properties}. In contrast, the previously studied amorphous polymers lack crystalline regions and therefore only soften above their glass transition temperature, before thermally degrading\cite{PlasticsTable}. The thermal melting drives the LDPE into a low-adhesion state, essential for successful transfer. The large difference in adhesion regimes offered by the LDPE phase transition is responsible for the exceptionally high success rate obtained with this method.


We observe that the process of monolayer pickup using hBN, outlined in ref.\cite{vdW_Pickup}, is highly dependent on the environmental relative humidity (RH), with both low (below 20\%) and high (around 60\%) RH values resulting in unreliable pickup. Here, the low surface energy of the bare LDPE, along with the possibility to increase it through plasma activation\cite{LDPE_Plasma}, provides a strong advantage by enabling careful tuning of the adhesion to match the necessary requirements. We verify that the plasma activation step along with covering monolayers with a 1-decanol SAM provides a robust protection against RH variations, making the pickup of hBN/monolayer structures a reliable and crack-free process, essential for optoelectronic devices. 

While we focused on transfer of GAE monolayers due to their intrinsic high quality and large area, not all 2D materials can be exfoliated using this technique. Notable examples of widely used 2D materials that cannot be easily exfoliated via GAE are graphene and hBN \cite{AuExfol}. We verify that the proposed transfer technique also works on large-area CVD-grown materials, showing similar increase in PL emission as seen in Fig. \ref{fig:1LTransfer_Characterization}b (see SI section 7).

Finally, we emphasize that while the force sensors employed in our transfer stage offer quantitative insights in the forces at play during the stamping process, the fundamental process does not rely on these sensors. As such, a simple motorized translation stage combined with an inexpensive temperature controller for the sample holder will readily enable the high-yield transfer of large-area monolayers and heterostructures.

\section{Conclusion}

In summary, we present a new and simple transfer method that capitalizes on the temperature-dependent adhesion and viscosity of LDPE to enable the transfer of large-area 2D material flakes, including GAE monolayer 2D materials and their heterostructures with hBN, onto substrates with arbitrary surface textures. We show the transfer of both large monolayers as well as  hBN/monolayer heterostructures onto patterned metasurfaces and sapphire domes to illustrate that even the most extreme surface textures that offer minimal adhesion can serve as the target substrate. This is in stark contrast to traditional methods which strongly rely on a flat surface with sufficient adhesion.
We demonstrate the application of the proposed method in combination with several patterned substrates to enable novel optical functionalities, including tailored angular emission patterns as well as strain- and electrostatic modulation of over excitonic emission. By fabricating large-area heterobilayers, we additionally confirm electronic coupling of the layers through the observation of interlayer exciton emission. We envision that this transfer method opens new avenues for the integration of large area (monolayer) 2D materials with nanostructured surfaces, enabling the exploration of new 2D materials physics and the development of novel optoelectronic devices that leverage the unique properties of atomically thin 2D semiconductors.

\section{Methods}
\subsection{Exfoliation of 2D materials}
Using the Au assisted exfoliation method reported previously\cite{AuExfol_Science}, we obtain large-area monolayers for transfer. A 100\,nm Au layer is deposited on a cleaned Si wafer at a rate of 0.5 Å/s, using electron beam physical vapor deposition (Polyteknik Flextura M508 E). A 300\,nm layer of PMMA is then spin coated on top of the Au to provide more mechanical stability during the peeling process. We use thermal release tape (TRT, Revalpha RAY-4LSC(N), Nitto Denko Corporation) and pick up the PMMA/Au layers, after which we press it immediately against a bulk TMDC crystal (WS$_2$ and WSe$_2$, HQ Graphene). After the exfoliation, we transfer the monolayer to cleaned SiO$_2$ substrates. The substrates are then heated to 110°C to remove the TRT, cleaned with acetone to remove the PMMA and placed in Au etchant (651818, Sigma–Aldrich) for 2 minutes to remove the gold. Finally, the sample is washed in isopropanol and dried with a nitrogen gun. 

hBN flakes are manually exfoliated from the bulk crystal (HQ Graphene)  using Nitto SPV-224 tape and transferred to a Si substrate using TRT. 

For the gated device, commercial large-area graphene monolayer is used (CVSO1011, ACS Material).

For fabrication of hBN/monolayer heterostructures, the large-area monolayers are covered in 1-decanol following the procedure outlined in ref.\cite{DodecanolCovering}. The substrates with monolayers are immersed in 1-decanol (8034631000, Sigma-Aldrich) for 2 minutes at 160°C. After this, the substrates are rinsed in IPA and dried using N$_2$. 

\subsection{Fabrication of the stamps}

The stamp is fabricated starting with a glass slide covered with two layers of PDMS with different thicknesses, cut from Gel Pak WF-40×40-0060-X0-A and AD-22T-00-X4. On top of the PDMS, a drop of heat-resistant superglue is placed and left to dry overnight. A square of LDPE film (commercially available as kitchen cling film, ``G'woon vershoudfolie'') is then cut from the cling film and stretched over the stamp, so that it conforms to the superglue drop. The LDPE film is held onto the stamp using Scotch tape. Details and illustrations on the stamp fabrication are provided in section 1 of the SI. For fabrication of hBN / monolayer heterostructures, the fabricated stamps are placed in an air plasma chamber at 100 W for 2 minutes before use.

\subsection{Stamping setup and force measurements}

For stamping of the 2D material flakes, a custom-built stamping microscope is developed. The substrate with GAE monolayer is placed on a heater that allows temperature control. The stamp is manipulated by a $xyz$ motorized platform (MT1-Z9, Thorlabs), allowing full control over direction and velocity of the motion. The process can be imaged through the stamp using either a Nikon Plan Fluor 4$\times$ (NA = 0.13) or 10$\times$ (NA = 0.3). 

The stamping tool is equipped with six load cells (SparkFun 14727) incorporating a Wheatstone bridge configuration to detect changes induced by mechanical strain. The signal is read using an HX711 load cell amplifier, performing analog-to-digital conversion. The amplifier is then interfaced with a microcontroller for real-time force monitoring and control.

\subsection{Spectroscopic measurements}

Room temperature spectroscopic measurements are performed using a Witec $\alpha$300R confocal Raman microscope. Here, a Zeiss EC Epiplan Neofluar 100$\times$ BD (NA = 0.9) is used with a 532\,nm laser as the light source. For the BFP measurements, a Zeiss Plan-Apochromat 100$\times$ (NA = 1.4) oil-immersion objective is used to image beyond the light line in air. 

For the wide field PL images a 405\,nm laser focused on the back focal plane of the objective is used, providing wide field near-normal illumination of the sample. The laser light is then filtered by a 600\,nm long-pass filter, allowing real space imaging of the sample's PL.

Measurements at cryogenic temperatures are performed using a cystom-built optical setup and a Montana Cryoadvance 50 cryostat, using an in-vacuum Zeiss EC Epiplan-Neofluar 0.9NA 100$\times$ objective for PL measurements. 

\subsection{Electrical connections and gating measurements}

To enable electrical contacts during optical characterization, the sample is mounted onto a custom printed-circuit board (PCB) using double-sided tape. The gate and ground electrodes are wire-bonded to separate contact pads on the PCB using 25 \textmu m diameter aluminum wires (West Bond 7KE). For DC gating measurements, a gate bias is applied using a Keithley 2612B SourceMeter. Initially, the sample is tested to establish the limits of maximum PL enhancement (+4\,V) and full exciton quenching (-8\,V). The voltage was subsequently swept from +4 to -8\,V in 1\,V increments. The experiment is fully automated using custom Python scripts to control the equipment.

\subsection{Device Characterization}

For the AFM measurements, a Bruker Dimension FastScan system is used at a rate of 5 \textmu m/s.  The SEM images are taken in a Thermo Fisher Scientific Helios UX 5 system. For the hBN/WS$_2$ on sapphire domes, the sample is first coated with a 5\,nm Au layer using a sputtering tool to prevent charge accumulation.

\subsection{Patterned Surfaces}

The hyperuniform pattern is fabricated on a cleaned Si substrate using electron beam lithography (Raith Voyager lithography system) to pattern the nanoholes in a CSAR resist layer. After development, the nanoholes are etched in the Si using Cl$_2$ + HBr/O$_2$ plasma at a rate of 3.3\,nm/s (Oxford PlasmaPro100 Cobra). Lastly, a SiO$_2$ layer is grown on the sample using rapid thermal annealing (AnnealSys AS-one 100) at 1100°C under O$_2$ atmosphere (50 sccm). The substrate with sapphire domes is purchased from MSE Supplies (WA0433).

\begin{acknowledgement}
The authors thank Paul Hanrahan and Thomas J. Kempa for supplying CVD-grown WS$_2$.
This work was funded by the Open Technology Program of the Dutch National Science Foundation (NWO), grant number 19486. JvdG is also supported by a Vidi grant (VI.Vidi.203.027) from the Netherlands Organization for Scientific Research (NWO), as well as an European Research Council Starting Grant under grant agreement No. 101116984. 
\end{acknowledgement}

\begin{suppinfo}
\begin{itemize}
    \item Description of fabrication of stamps for 2D material transfer.
    \item Description of the stamping tool and force measurements.
    \item Detailed spectral analysis of the PL and Raman spectra at multiple transfer steps. 
    \item Calculation of the structure factor for the hyperuniform metasurface.
    \item AFM measurement of hBN flake thickness before and after plasma etching.
    \item Demonstration of transfer of CVD grown monolayers.
    \\
\end{itemize}

\textbf{Data availability} - A full replication package including all data and analysis scripts will be made freely available upon publication.

\end{suppinfo}


\newpage
\section{Supplementary information}
\section{LDPE Stamp Fabrication}

\begin{figure}[!ht]
    \centering
    \includegraphics[width=1\linewidth]{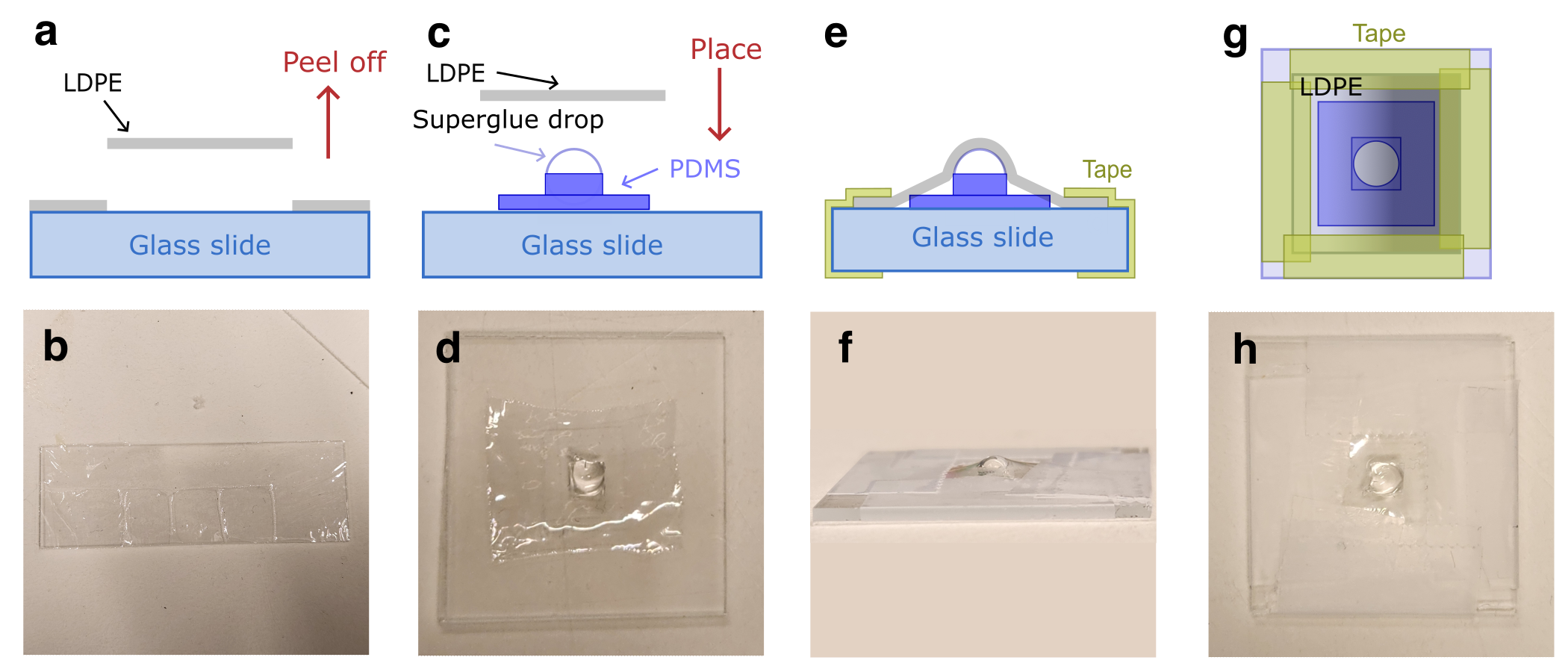}
    \caption{\textbf{Fabrication of the stamps for 2D material transfer.} \textbf{(a,c,e,f)} Schematics of the different steps of stamp fabrication. \textbf{(b,d,f,h)} Photographs of the stamp at each step of the fabrication.}
    \label{SIFig1}
\end{figure}

Here we detail the fabrication of the stamps used for the transfer of monolayer and hBN / monolayer heterostructures. The LDPE is stretched and placed on a glass slide, and cut in square shape with a sharp knife, (Fig. S\ref{SIFig1}a-b). A glass slide is then prepared with a double PDMS layer, with the bottom layer cut from Gel Pak WF-40×40-0060-X0-A and the top one from AD-22T-00-X4 (Fig. S\ref{SIFig1}c). These layers ensure good adhesion of the whole structure to the glass and create an offset distance to prevent the stamp from touching the 2D materials anywhere else except from the dome top. A drop of superglue is then manually placed on top of the thicker PDMS and left overnight to dry. The PDMS-superglue stack also creates a soft set of polymers that allows the stamp to conform to the 2D material without applying too much pressure, which could lead to cracks in the transfer. The LDPE layer is then placed on the structure (Fig. S\ref{SIFig1}d) and streched with tape, ensuring good contact between the LDPE and superglue (Fig. S\ref{SIFig1}e-f). Figures S\ref{SIFig1}g and h demonstrate the final result from stamp fabrication. 

\begin{figure}[!ht]
    \centering
    \includegraphics[width=1\linewidth]{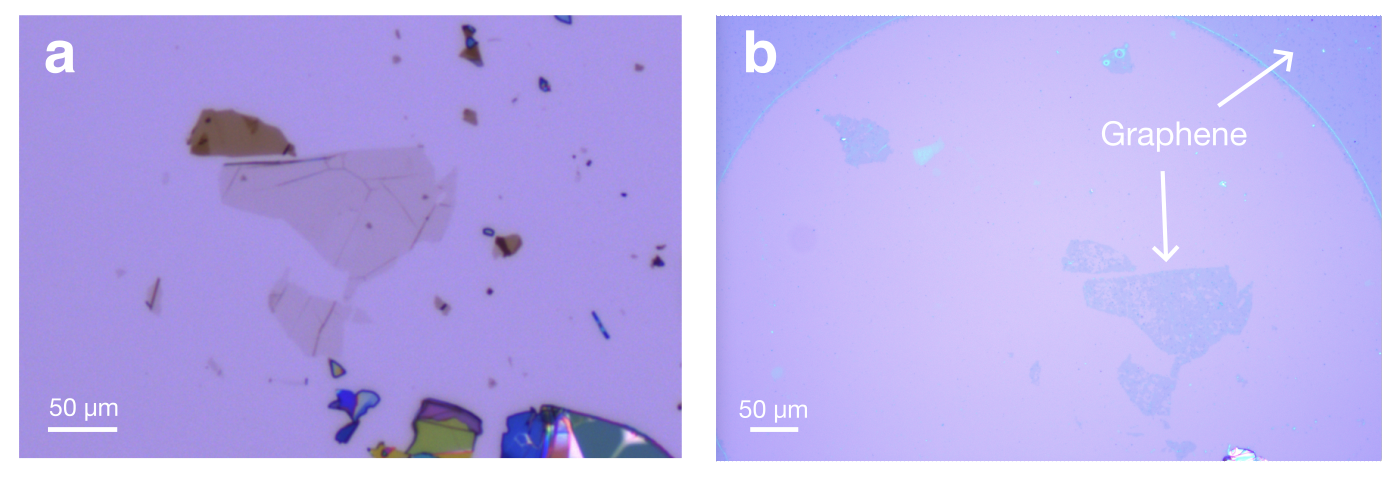}
    \caption{\textbf{Attempt to pick up monolayer graphene with hBN without plasma and decanol treatment.} \textbf{(a)} hBN flakes used in the attempt. \textbf{(b)} Graphene layer after pickup attempt, where the shape of the hBN flakes is clearly visible.}
    \label{SIFig2}
\end{figure}

As mentioned in the main text, we perform an air plasma activation of the LDPE after stamp fabrication and additionally create a decanol self assembled monolayer on the 2D material monolayer, as described in the main text. Without this step, we verify that the pickup of monolayers under hBN is not successful, possibly due to surface energy mismatch between the materials and the stamp, leading to trapping of interfacial water or contaminants. Figures S\ref{SIFig2}a and b show an attempt to pick up graphene with hBN. We verify that after the procedure described in the main text, a large portion of the graphene was picked up (where it was contacting with LDPE), exposing the SiO$_2$/Si substrate. Nevertheless, in the regions where the hBN flakes were present, we see a perfect outline of the flake shape, indicating that the graphene remained in the substrate and that the pickup failed. 

\section{Stamping tool and force measurements}

\begin{figure}[!ht]
    \centering
    \includegraphics[width=.5\linewidth]{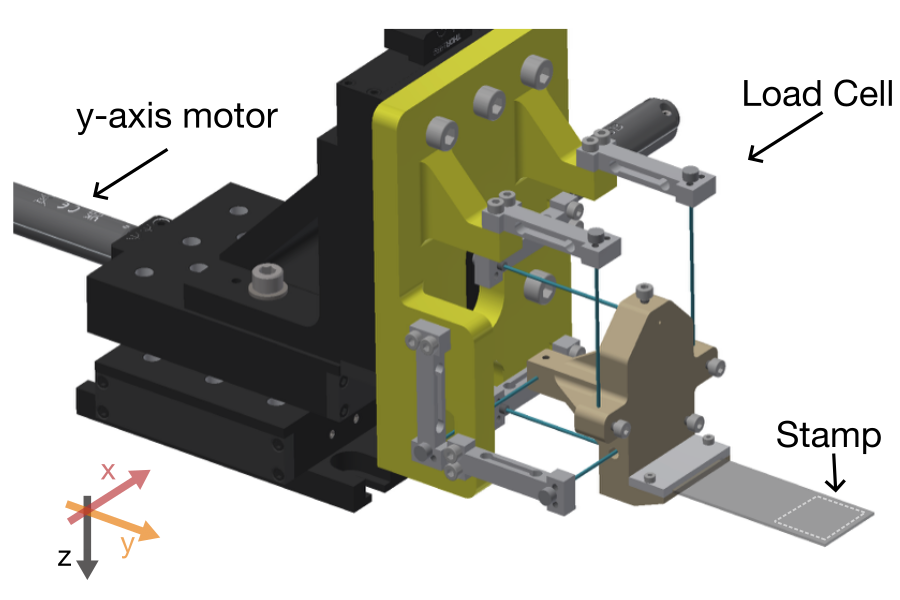}
    \caption{\textbf{Schematic of the stamping tool equipped with load cells for force measurement of 2D material transfer.}}
    \label{SIFig3}
\end{figure}

Stamping is performed using a motorized stamping tool in $x$, $y$ and $z$ directions, allowing precise control of the contact with the monolayer (Fig. S\ref{SIFig3}). 
The stamping tool includes six load cells that hold the stamp and allow for independent measurement of the forces along the three axes. The stamp is suspended along the $y$ axis, allowing contact with the monolayer without interference from the rest of the tool. It should be noted that under this geometry, for forces that induce some curvature on the glass slide, $F_y$ and $F_z$ are not fully independent. This coupling is caused by the deformation of the glass slide under because the beam’s deflected shape increases its arc length, requiring an axial support force to satisfy the glass inextensibility, so a change in $F_z$ inherently leads to a generation of a force in the $y$ direction (axial) \cite{Deformation}.

\section{Spectral analysis of the transferred monolayer}

\begin{figure}[!ht]
    \centering
    \includegraphics[width=.5\linewidth]{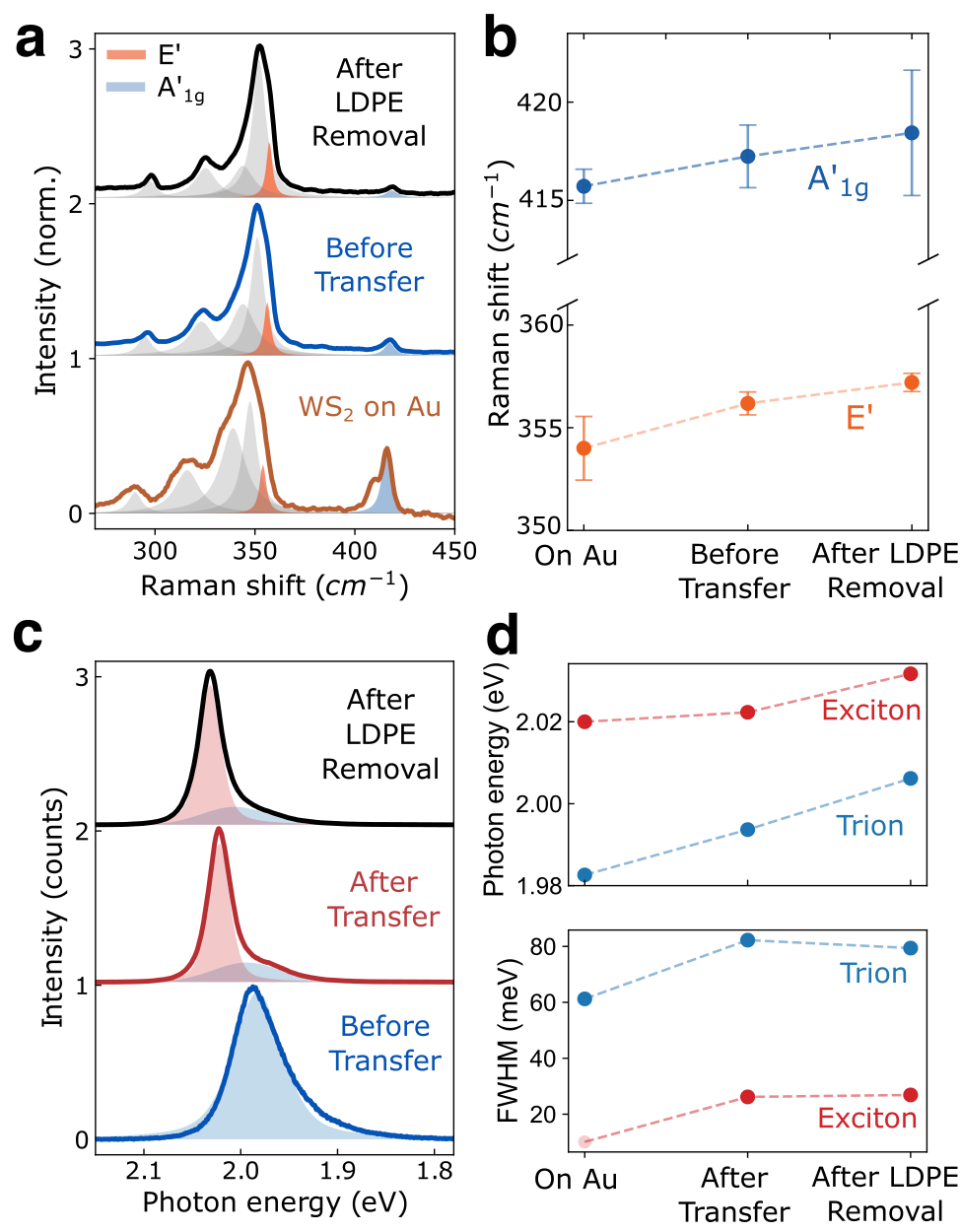}
    \caption{\textbf{Analysis of the monolayer WS$_2$ Raman and PL spectra in the different steps of transfer.} \textbf{(a)} Raman spectra with the respective Lorentzian line fits. \textbf{(b)} Raman shift of the A'$_{1g}$ and E' peaks at different steps of the transfer. \textbf{(c)} PL spectra with the respective Lorentzian line fits. \textbf{(d)} Neutral exciton and trion peak energies and FWHM at different steps of the transfer.}
    \label{SI_SpectraPL}
\end{figure}

To explain the change in the PL spectra between the different transfer steps, we turn to a deeper analysis of the spectral lineshape and peak positions (Fig. S\ref{SI_SpectraPL}). 

From Fig. 2b of the manuscript, it is apparent that the largest change occurs between the as-exfoliated sample on a SiO$_2$ substrate and its subsequent pickup and transfer to a new SiO$_2$ substrate, still with LDPE residue on top. On that path, the radiative rate of neutral excitons, assigned to the intense peak around 2.02-2.03 eV emerges and vastly exceeds that of the trion peak around 1.98-2.0 eV. Such a change is characteristic of a large change in the strain or Fermi level in the monolayer.

First, we address possible contributions from strain by fitting Lorentzian contributions at the known WS$_2$ peak positions to the Raman spectra at different steps of the transfer. Comparing the Raman spectra of the WS$_{2}$ monolayer on the gold pickup to the same monolayer on a SiO$_2$ substrate after the GAE is complete, we see a hardening by 2.2 cm$^{-1}$ of the E’ peak around 354 cm$^{-1}$, which is an in-plane vibration and most sensitive to strain (Fig. S\ref{SI_SpectraPL}a,b).\cite{Michail2023, Negi2025} While there is a further hardening of this mode upon stamping and transfer with LDPE to a new SiO$_2$ substrate, this shift is more subtle, only 1 cm$^{-1}$. On the other hand, the A’$_1$ mode (416 cm$^{-1}$) continues to shift to higher wavenumbers throughout the process, indicating a continued reduction in the doping level of the monolayer.\cite{Sohier2019} We therefore argue that a change in the strain is unlikely to be the origin of the shift from trion-dominated to neutral exciton-dominated PL upon transfer. 

The changes to the fitted peak positions of the neutral exciton and especially trion contributions also point to doping as the largest factor in changing the PL signature (Fig. S\ref{SI_SpectraPL}c,d). Specifically the continued blue shift in the trion peak throughout the process is indicative of a continued change in the Fermi level.\cite{Mak2013} 

There are further competing mechanisms when assigning the origin of the shift in doping level of the monolayer through the stamping process. We rule out contributions from the substrate, as both the initial and targets are SiO$_2$, cleaned in the same way and not exposed to high temperatures and aggressive environments which are known to alter its surface chemistry. This leaves solvents from the removal of the PMMA and Au from the initial exfoliation, effects from the plasma-activated functional groups on the LDPE stamp surface, and passivation of defects and charge neutralization by oleic acid, which we use to remove the LDPE stamp. Because the largest change comes upon removal from the initial substrate, we argue that the intercalated polar solvents are likely the cause of the initial doping and thus trion-dominated PL. Poddar et al. have shown that exposure to acetone and IPA cause unintentional doping seen by large shifts to the threshold voltage of monolayer Mo$_2$ transistors \cite{Poddar}. 

We cannot, however, decouple the possible effects of charge transfer from plasma-activated functional groups on the LDPE stamp. We use air plasma to activate the LDPE surface, which is known to create carbonyl groups, which are electron withdrawing.\cite{Sanchis2006, Pandiyaraj2014} As WS$_2$ monolayers are known to be slightly n-doped due to the presence of chalcogen vacancies, the interaction between the LDPE stamp and the monolayer may also induce a reduction in the Fermi level through charge transfer.\cite{Shokouh2015, Ma2018, McGinn2024} This we also cannot decouple from the affects of the oleic acid, with which we remove the LDPE stamp, which is also known to passivate chalcogen vacancies and neutralize MoS$_2$ and WS$_2$. Taken together, these three mechanisms explain the large spectral changes we observe in PL of the WS$_2$ monolayer throughout the stamping process. 

\section{Calculation of the structure factor}

In the hyperuniform pattern presented in the main text, the structure factor is calculated using images obtain from a SEM. 

\begin{figure}[!ht]
    \centering
    \includegraphics[width=.7\linewidth]{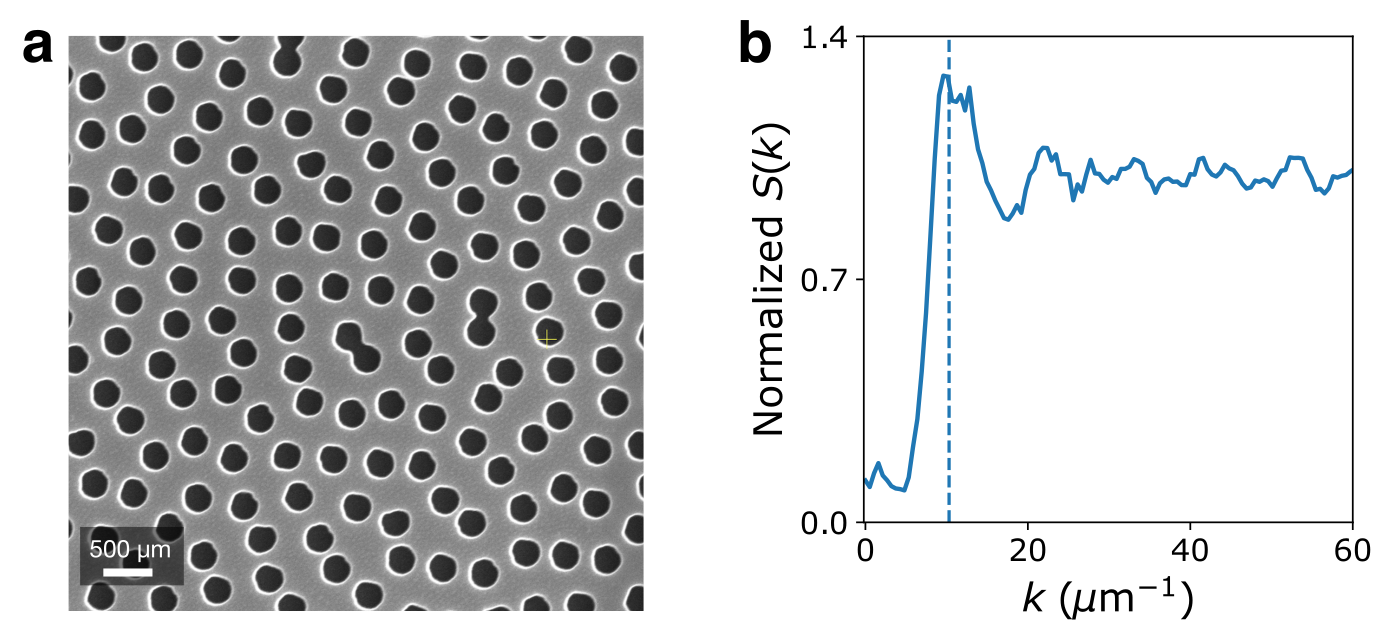}
    \caption{\textbf{Calculation of the structure factor S($k$) of a hyperuniform pattern.} \textbf{(a)} SEM image used for calculation of S($k$). \textbf{(b)} Calculated S($k$). The dashed line corresponds to the maximum value registered in the BFP PL image of the device.  }
    \label{SIFig5}
\end{figure}

Taking Fig. S\ref{SIFig5}a, the structure factor S($k$) was calculated using Fiji \cite{Fiji} to threshold and extract the position of the center of each nanohole. To calculate S($k$), we used\cite{Hyperuniform} 

$$S(\mathbf{k}) = \frac{1}{N} \left\langle \sum_{k,j=1}^N e^{-i\mathbf{k}\cdot(\mathbf{R}_j - \mathbf{R}_k)} \right\rangle,
$$

where $\mathbf{k}$ is the wavevector, $N$ is the total number of nanoholes and $\mathbf{R}_{j,k}$ are the positions of each nanohole. The maximum of $S(\mathbf{k})$ is obtained at $k$ = 10.3 $\mu$m$^{-1}$, in agreement with the BFP measurements shown in the main text, that demonstrate PL enhancement at this value of the wavevector (Fig. S\ref{SIFig5}b). 

\section{Plasma removal of LDPE residue}

Using AFM measurements, we demonstrate that the air plasma approach does not induce any thickness change in the hBN flakes, allowing this method to be used in nanophotonic applications, where dielectric thickness can be crucial (Fig. S6). 

\begin{figure}[!ht]
    \centering
    \includegraphics[width=.7\linewidth]{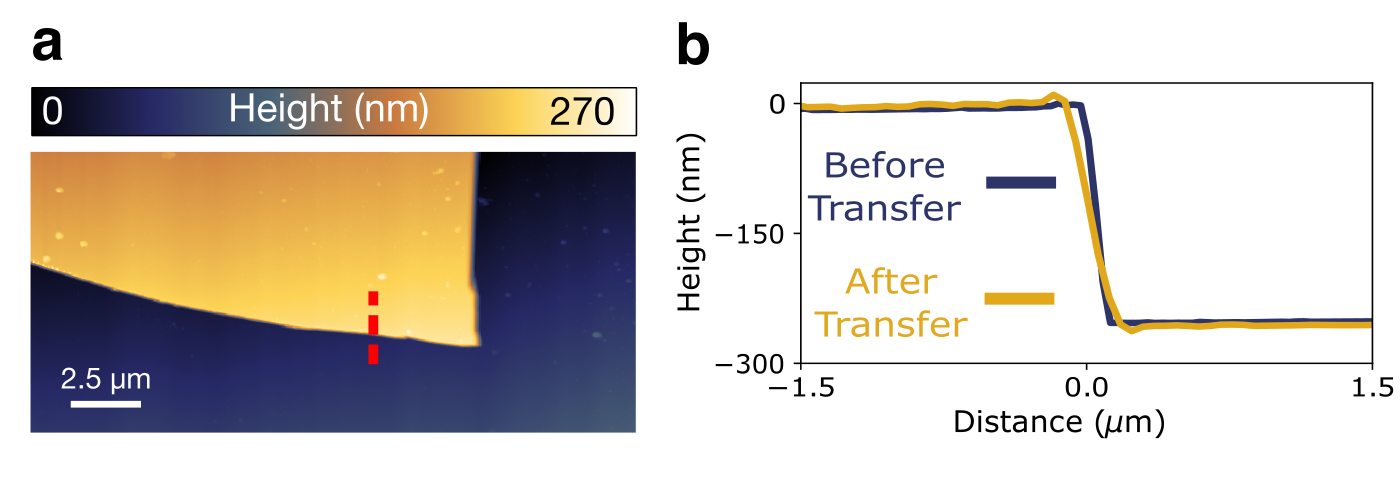}
    \caption{\textbf{AFM scan of the hBN flake used for monolayer transfer.} \textbf{(a)} Full AFM scan of the hBN flake before transfer. The red dashed line indicates the cross section used in (b). \textbf{(b)} Thickness measurement before and after air plasma removal of LDPE.}
    \label{SIFig6}
\end{figure}

We verify that no significant vertical etching of the hBN flake is observed, only a slight tilt of the hBN wall can be obtained for prolonged plasma exposure, this effect is not significant for most applications because the hBN lateral size is much larger than the scale of this effect.

\section{Transfer of CVD-grown monolayers}

In the manuscript we focused on transfer of GAE monolayers due to their inherent high-quality and large areas, enabling combination with large patterned surfaces. While applicable in many scenarios, not all materials can be exfoliated using this technique, and CVD growth remains one of the most common methods to obtain large area monolayers. Here, we demonstrate that the same monolayer transfer method used in the manuscript applies also for CVD grown materials.

\begin{figure}[!ht]
    \centering
    \includegraphics[width=1\linewidth]{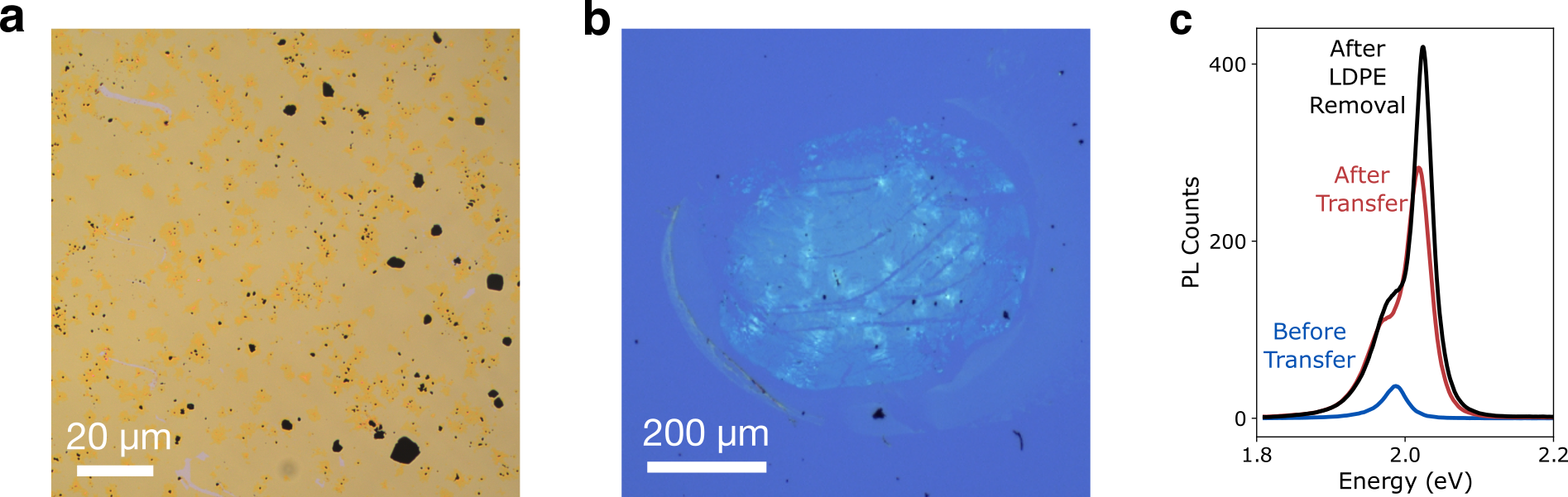}
    \caption{\textbf{Transfer of CVD-grown WS$_2$ monolayer.} \textbf{(a)} Bright field image of the CVD-grown WS$_2$ monolayer, before pick-up. The darker yellow patches correspond to bilayer growth, while the dark spots correspond to the seed crystals used for monolayer growth. \textbf{(b)} Bright field image of the transferred WS$_2$ monolayer. \textbf{(c)} PL spectra of the monolayer before and after transfer, as well as after LDPE residue removal.}
    \label{SIFig7}
\end{figure}

CVD-grown WS$_2$ on SiO$_2$ (300 nm) / Si was transferred to a substrate of the same materials, following the recipe discussed in the manuscript. Overall, we observe similar results to the GAE transfer, with full monolayer and bilayer pickup and transfer to the target substrate (Fig. S7a,b). Analysis of the PL (Fig. S7c) demonstrates a similar trend after transfer, with a 12x increase of the PL intensity followed by a linewidth reduction from 40 meV (before transfer) to 27 meV (after LDPE removal). Here, a marked difference to the GAE-exfoliated monolayer after transfer is the much stronger trionic contribution, indicative of possible defects caused by the growth process of the CVD monolayer. 

\bibliography{achemso-demo}

\end{document}